\documentclass[10pt,twocolumn,a4paper]{article}
\usepackage[inner=0.70in,outer=0.70in,top=0.68in,bottom=0.67in,columnsep=0.25in]{geometry}
\usepackage[T1]{fontenc}
\usepackage[utf8]{inputenc}
\usepackage{mathptmx}
\usepackage{courier}
\usepackage{amsmath,amssymb,graphicx,booktabs,multirow}
\usepackage[table]{xcolor}
\usepackage[numbers,super,sort&compress]{natbib}
\usepackage{stfloats}
\usepackage{flushend}
\usepackage{titlesec}
\usepackage[font=small,labelfont=bf,skip=4pt]{caption}
\usepackage{textcomp}
\usepackage{pdfpages}
\usepackage[hidelinks]{hyperref}
\usepackage{microtype}
\titleformat{\section}{\large\bfseries}{\thesection}{0.6em}{}
\titleformat{\subsection}{\normalsize\bfseries}{\thesubsection}{0.55em}{}
\titlespacing*{\section}{0pt}{12pt plus 2pt minus 1pt}{5pt}
\titlespacing*{\subsection}{0pt}{8pt plus 2pt minus 1pt}{3pt}
\hypersetup{
  pdftitle={DREChanNet: Patient-Relative iEEG Model for Channel-Wise Epileptogenic Zone Localization},
  pdfauthor={Yuanchu Gong, Zibo Yan, Yibo Lyu, Zhe Shi, Qiang Li, Yuanfeng Zhou, Minqiang Yang, Yalin Wang, Wei Chen, Bin Hu}
}
\begin{document}
\raggedbottom
\raggedend
\twocolumn[{%
  \centering
  {\fontsize{17}{20}\selectfont\bfseries
  DREChanNet: Patient-Relative iEEG Model for\\
  Channel-Wise Epileptogenic Zone Localization\par}
  \vspace{7pt}
  {\normalsize Yuanchu Gong\textsuperscript{1}, Zibo Yan\textsuperscript{1},
  Yibo Lyu\textsuperscript{1}, Zhe Shi\textsuperscript{2}, Qiang Li\textsuperscript{2},\\
  Yuanfeng Zhou\textsuperscript{3}, Minqiang Yang\textsuperscript{1,*},
  Yalin Wang\textsuperscript{1,*}, Wei Chen\textsuperscript{4}, and Bin Hu\textsuperscript{1,*}\par}
  \vspace{5pt}
  {\small
  \textsuperscript{1}\,School of Information Science and Engineering, Lanzhou University, Lanzhou 730000, China\\
  \textsuperscript{2}\,The Second Hospital of Lanzhou University, Lanzhou 730030, China\\
  \textsuperscript{3}\,Children's Hospital of Fudan University, Shanghai 201102, China\\
  \textsuperscript{4}\,School of Biomedical Engineering, University of Sydney, Australia\par}
  \vspace{4pt}
  {\footnotesize
  \textsuperscript{*}\,Correspondence: Minqiang Yang (\href{mailto:yangmq@lzu.edu.cn}{yangmq@lzu.edu.cn}),
  Yalin Wang (\href{mailto:yalinwang@lzu.edu.cn}{yalinwang@lzu.edu.cn}),
  and Bin Hu (\href{mailto:bh@lzu.edu.cn}{bh@lzu.edu.cn})\par}
  \vspace{10pt}
  \begin{minipage}{\textwidth}
  \small\noindent\textbf{Abstract.} Accurate localization of the epileptogenic zone (EZ) is essential for presurgical evaluation of patients with drug-resistant focal epilepsy. Intracranial electroencephalography (iEEG) provides direct recordings of seizure-related brain activity, yet identifying epileptogenic contacts remains challenging because background electrophysiological activity and electrode coverage vary substantially across patients. These differences complicate the interpretation of absolute signal measurements and highlight the need to account for patient-specific recording context. We propose a patient-relative (PR) plug-in that expresses physiological features or learned representations relative to the distribution of recorded channels within each patient. This label-free operation can be incorporated as an input transformation or a lightweight correction to an existing predictor. We further introduce DREChanNet, a compact ictal localization model integrating six physiological descriptors with learned raw-signal representations. Same-channel pre-onset contrasts characterize seizure-related changes, cross-seizure aggregation preserves response magnitude and variability, and PR contextualizes both branches before prediction. We investigate within-patient physiological associations and held-out-channel reference prediction, followed by baseline comparisons, component ablations, cross-source transfer, seizure-sufficiency analysis, and channel ranking. Evaluation includes an ictal cohort of 80 patients and a patient-based interictal cohort of 124 patients. PR improves macro-F1 by an average of 9.03 percentage points across eight ictal baselines and improves the separately fitted interictal readout. These findings support patient-relative contextualization as a reusable approach to channel-level localization and establish DREChanNet as an integrated model for ictal iEEG analysis.

  \vspace{4pt}
  \end{minipage}\par
  \vspace{5pt}
}]

\section{Introduction}
Drug-resistant epilepsy (DRE) creates a localization problem with direct consequences for treatment. Surgery can provide seizure freedom for appropriately selected patients after medication failure \cite{ref1,ref2,ref3}, and international recommendations support timely surgical evaluation \cite{ref4}. The central decision is which tissue to treat while preserving neurological function (Fig. 1). Intracranial electroencephalography (iEEG) informs this decision through measurements at implanted contacts \cite{ref5,ref6}.

\begin{figure}[!htb]
\centering
\includegraphics[width=\linewidth]{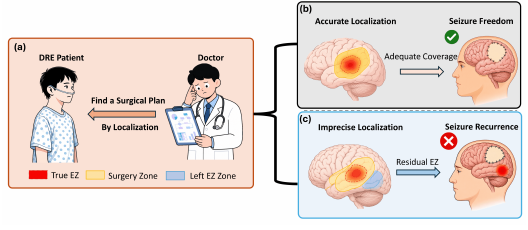}
\caption{Clinical motivation for EZ localization. (a) Presurgical assessment identifies a treatment target. (b) Coverage of the seizure-generating tissue supports seizure freedom. (c) Residual epileptogenic tissue can sustain seizures. The surgical region and the underlying epileptogenic target need not have identical boundaries.}
\label{fig:1}
\end{figure}

Localization connects a tissue-level treatment target to selectively sampled electrophysiology. The epileptogenic zone (EZ) denotes tissue whose treatment is required for seizure freedom; iEEG measures activity at locations selected to test a presurgical hypothesis \cite{ref5,ref7}. Coverage therefore reflects clinical suspicion and implantation constraints. Human microelectrode recordings further show that large ictal field potentials can extend beyond the territory of intense neuronal recruitment \cite{ref8}. The seizure onset zone (SOZ), early propagation, and treatment volume describe different stages of localization. A recent international survey documented differing clinical definitions of the SOZ, while a Delphi study established confidence grades for EZ evidence \cite{ref9,ref10}. A channel score therefore belongs to a clinical inference task conditioned on observed brain regions, recording state, and clinical annotations.

Quantitative approaches extract complementary localization information. Ictal indices combine fast activity with recruitment timing \cite{ref11}; interictal approaches characterize waveform events, spectral changes, and network dynamics \cite{ref12,ref13,ref14}; and neural models learn representations from the recorded signals \cite{ref15,ref16}. Recent syntheses identify spatial sampling, label definitions, and validation across patients as central challenges in translating such measurements into clinical decisions \cite{ref17,ref18,ref19}. Together, these studies motivate a question that cuts across the choice of biomarker or encoder: against which reference should a new patient's channel be interpreted?

The reference matters because physiological variation is structured. An atlas of normal iEEG demonstrates substantial regional differences in spectral activity \cite{ref20}, and anatomical normative mapping uses this variation to interpret patient abnormalities \cite{ref21,ref22}. Recording time also affects the stability of channel abnormality maps \cite{ref23}. Thus, a channel value carries information about its biological and recording context as well as its relationship to the clinical target. Cross-patient learning must preserve useful within-patient differences while accommodating the baseline distributions associated with each new implantation. Alongside an external population reference, another source of context is already available when a patient is assessed: the distribution of that patient's observable channels. It indicates whether a coordinate is relatively elevated or reduced within the implanted set, without requiring the channel labels at inference.

We make this patient-relative reference an explicit, reusable part of localization. The proposed patient-relative (PR) plug-in standardizes each feature or embedding coordinate across the patient's channels. In a linear readout, this transformation produces patient-dependent coordinate weights and an offset, allowing the same learned predictor to interpret different patients through their own reference statistics. PR can replace the physiological feature inputs of an existing classifier or supply a learned correction to a frozen model's output. This separates the reference operation from the architecture used to encode the signal and makes its contribution testable across models.

We further develop DREChanNet, a compact ictal model that combines physiological and raw-signal evidence. Six physiological descriptors provide interpretable measurements, while a temporal encoder captures waveform structure. Same-channel pre-onset subtraction represents temporal change, and cross-seizure means and standard deviations retain recurrent responses and their variability. PR processing then expresses both branches relative to the patient's channel set before a shared head combines them. The model compares a channel with its earlier activity and with other channels in the patient.

The evaluation follows this sequence of reasoning. Within-patient feature associations establish the relevance of the descriptors; held-out-channel distribution prediction tests the reference itself; and paired baseline comparisons and component ablations assess its contribution to localization. Patient-held-out evaluation on the 80-patient ictal cohort examines source transfer, seizure availability, and ranking. A 124-patient interictal evaluation derived from Omni-iEEG \cite{ref24} tests PR with an existing backbone. The contributions are:

\begin{itemize}
\item A PR plug-in that makes the observed patient channel distribution available to feature classifiers and learned representations, supported by feature-association and reference-prediction analyses.
\item DREChanNet, a 9,329-parameter ictal model integrating physiological and temporal representations through pre-onset referencing, cross-seizure summaries, and PR processing.
\item A transparent, reproducible evaluation pipeline integrating the most comprehensive accessible ictal and interictal cohorts under strict patient-based definitions, establishing a verifiable benchmark across classification, ranking, source transfer, and seizure availability.
\end{itemize}

\section{Related Work}
Quantitative iEEG localization combines signal and network evidence. The epileptogenicity index links fast activity to the timing of ictal recruitment \cite{ref11}. Human recordings distinguish recruited seizure cores from surrounding penumbral territories \cite{ref8}, with sustained ictal high-frequency activity informing that distinction \cite{ref25}. Interictal approaches combine epileptiform discharges, oscillatory events, and cross-frequency coupling \cite{ref12}, with useful localization information also demonstrated in routinely sampled frequencies below 45 Hz \cite{ref26}. Dynamical perturbations, directed connectivity, and network suppression characterize channel participation in the recorded system \cite{ref13,ref14}, \cite{ref27,ref28}. Together, these studies motivate complementary physiological and temporal representations, whose contributions we examine through association and ablation analyses.

Reference distributions and learned encoders offer complementary ways to organize this evidence. The normal iEEG atlas identifies region-dependent spectral structure \cite{ref20}, and normative mapping relates deviations in activity or connectivity to epileptogenic tissue \cite{ref21,ref22}. Their reference distributions are indexed by anatomy. Temporal analyses assess their stability across recording segments \cite{ref23}. In parallel, SEEG-Net uses multiscale convolutions and domain-generalization training for cross-subject pathological activity detection \cite{ref15}, and multiscale neural models have localized seizure-onset channels from brief interictal recordings \cite{ref16}. SEEGformer combines signal representations, anatomical information, and cross-channel attention in an individual-specific detection and localization model \cite{ref29}. PR makes a different reference directly available to these representations: the distribution of observable channels within the patient being assessed. This common operation can support joint training or an added readout.

Clinical interpretation depends on how methods are evaluated and how their targets are defined. Reviews of quantitative iEEG and computational DRE research identify opportunities for clinical use alongside challenges in spatial sampling, label construction, and validation \cite{ref17,ref18,ref19}. An ILAE systematic review of connectivity-based outcome prediction further motivates standardized reporting and patient-level data sharing \cite{ref30}. Channel-label agreement, outcome association, and prospective clinical benefit provide distinct forms of evidence. Shared resources support their investigation: iEEG-BIDS standardizes recording organization \cite{ref31}, while Omni-iEEG harmonizes recordings and metadata for multiple epilepsy tasks \cite{ref24}. Within these resources, the prediction unit remains consequential because windows and channels are nested within patients; evaluation must match the intended use on new patients \cite{ref32}. Our patient-held-out evaluation links reference quality to classification, ablation, and ranking, with explicit cohort construction and patient-equal summaries.

\begin{figure*}[!tb]
\centering
\includegraphics[width=\linewidth]{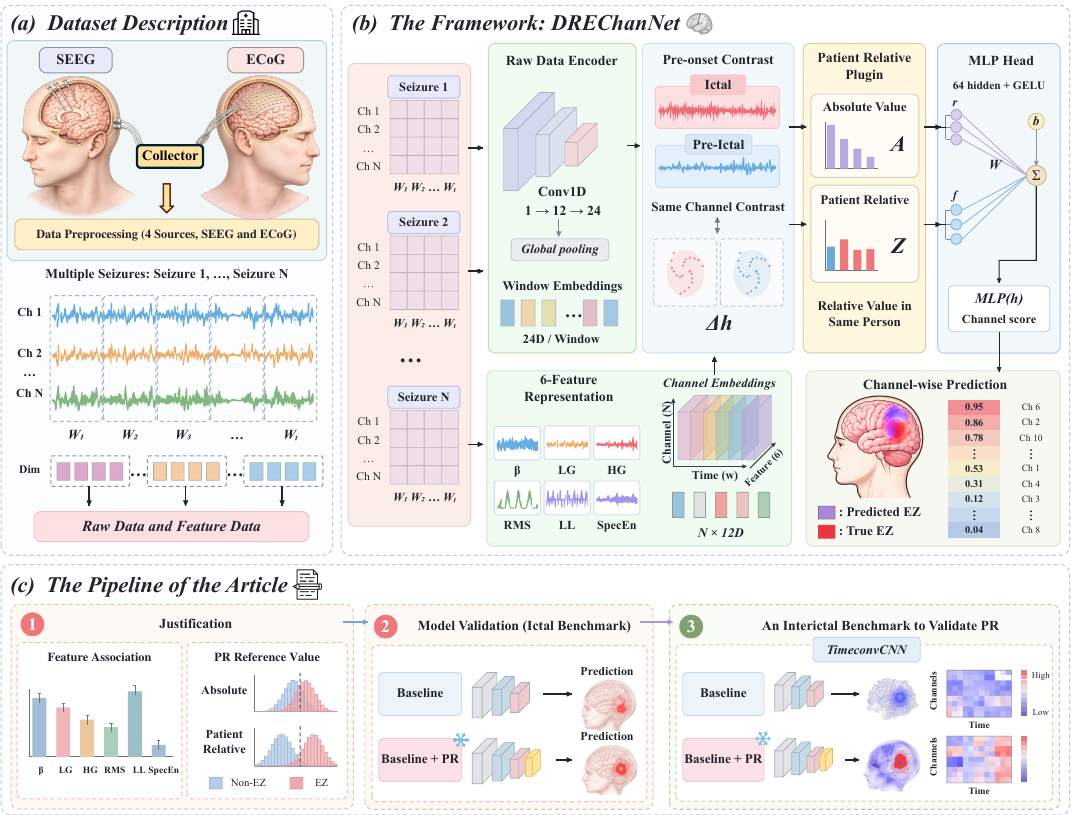}
\caption{DREChanNet combines physiological summaries and raw temporal representations. Each branch uses a patient-relative representation before a shared prediction head. Same-channel pre-onset subtraction defines temporal changes, whereas PR defines each channel's position within the patient's channel distribution. The exact aggregation and fusion are given in Eqs. (9)--(11).}
\label{fig:2}
\end{figure*}
\section{Methods}
\subsection{Task, Cohorts, and Patient-Level Evaluation}
For patient p, let $\mathcal C_p$ be the set of observable implanted channels and $y_{pc}\in\{0,1\}$ the clinical annotation of channel c, with EZ encoded as 1. The task is to assign an EZ score $q_{pc}$ and a binary decision to each channel. Our approach first characterizes physiological changes around seizure onset, then expresses channel representations relative to the patient's own channel set (Fig. 2). DREChanNet combines this patient-relative (PR) information in physiological and raw-signal branches; the same PR operator can also be attached to an existing predictor.

The ictal cohort includes 80 postoperative seizure-free patients, 256 seizures, and 7,635 channels, including 1,743 annotated EZ channels, from four data sources (Table 1). EZ/NEZ denotes source-specific clinical localization annotations, used as clinical target surrogates. The study received approval from the Second Hospital of Lanzhou University (2023A-765) and the Children's Hospital of Fudan University (2024-396).

Five outer folds contain 16, 16, 17, 15, and 16 patients. Each development partition is separated into fitting (FIT) and validation patients, with 13 validation patients per fold. All channels and recordings of a patient remain together. FIT patients determine learned parameters and population preprocessing, whereas validation patients determine checkpoints and decision thresholds. Outer patients are used for evaluation. Their observable unlabeled channels supply their own PR reference at inference.

\subsection{Physiological Channel Representations}
Each valid 2-s window sampled at 250 Hz provides six measurements: beta power (13--30 Hz), low-gamma power (30--80 Hz), partial high-gamma power (80--125 Hz), root-mean-square amplitude (RMS), line length (LL), and spectral entropy (SE; 10--125 Hz). With N = 500, window duration T = 2 s, and Welch spectrum S(f), they are computed as

\begin{equation}
\begin{aligned}P_{[a,b]}&=\log\left(1+\int_a^b S(f)\,df\right),\\
\mathrm{RMS}&=\sqrt{N^{-1}\sum_{n=1}^N x_n^2+10^{-8}},\\
\mathrm{LL}&=T^{-1}\sum_{n=2}^N|x_n-x_{n-1}|,\\
\mathrm{SE}&=-\frac{\sum_{m=1}^M\pi_m\log\pi_m}{\log M},\quad \pi_m=\frac{S(f_m)}{\sum_{r=1}^M S(f_r)}.\end{aligned}
\label{eq:1}
\end{equation}

\begin{table}[!htb]
\centering
\caption{Ictal localization cohorts. EZ/NEZ percentages refer to valid implanted channels.}
\label{tab:overall_cohort}
{\fontsize{8}{9.1}\selectfont
\setlength{\tabcolsep}{2.8pt}
\renewcommand{\arraystretch}{1.08}
\begin{tabular}{@{}lrrrr@{}}
\toprule
Cohort & Patients & Seizures & Channels & EZ/NEZ (\%) \\
\midrule
HUP~\cite{ref33}
& 36 & 117 & 3,667 & 16.8/83.2 \\

Multi-site~\cite{ref34}
& 15 & 45 & 925 & 17.7/82.3 \\

LZU
& 21 & 72 & 2,141 & 38.2/61.8 \\

Fudan
& 8 & 22 & 902 & 16.2/83.8 \\

\midrule
\textbf{Total}
& \textbf{80}
& \textbf{256}
& \textbf{7,635}
& \textbf{22.8/77.2} \\
\bottomrule
\end{tabular}
}
\end{table}

Welch estimation uses 500 samples and 0.5-Hz bins, numerical integration uses the trapezoidal rule, and the high-gamma band includes the available Nyquist bin. Entropy uses the M bins between 10 and 125 Hz, with 0 log 0 = 0.

Let $\psi_j(x_{pcsw})$ denote measurement j in window w of seizure record s. We first compute a same-channel pre-onset reference over valid pre-onset windows $\mathcal B_{pcs}$ and then average the resulting changes over all valid windows $\mathcal W_{pcs}$ in that seizure record:

\begin{equation}
\begin{aligned}b_{pcs,j}&=\frac1{|\mathcal B_{pcs}|}\sum_{w\in\mathcal B_{pcs}}\psi_j(x_{pcsw}),\\d_{pcs,j}&=\frac1{|\mathcal W_{pcs}|}\sum_{w\in\mathcal W_{pcs}}[\psi_j(x_{pcsw})-b_{pcs,j}].\end{aligned}
\label{eq:2}
\end{equation}
For the $S_{pc}$ available seizures, the mean and population SD are

\begin{equation}
\begin{aligned}m_{pc,j}&=\frac1{S_{pc}}\sum_s d_{pcs,j},\\v_{pc,j}&=\sqrt{\frac1{S_{pc}}\sum_s(d_{pcs,j}-m_{pc,j})^2}.\end{aligned}
\label{eq:3}
\end{equation}
Concatenating ($m_{pc,j}$, $v_{pc,j}$) for the six measurements yields the 12-dimensional physiological vector $f_{pc}$. Cross-seizure SD equals zero for the seven patients with one seizure. The temporal reference in Eq. (2) compares a channel with its own pre-onset activity; the PR reference below compares different channels within a patient.

We assess within-patient EZ association using the signed rank-biserial effect

\begin{equation}
D_{pj}=\frac{2U_{pj}}{n_{p,\mathrm{EZ}}n_{p,\mathrm{NEZ}}}-1,
\label{eq:4}
\end{equation}
where U is the Mann--Whitney statistic, with half credit for ties between EZ and NEZ values. Positive D indicates larger values in EZ channels. Patient effects receive equal weight; 10,000 patient-bootstrap resamples provide nominal 95\% intervals. Two-sided Wilcoxon signed-rank tests use Benjamini--Hochberg correction across the 12 coordinates. Source-specific means and proportions of positive effects describe their distribution across cohorts.

\subsection{The Patient-Relative Operator}
For any channel representation $z_{pc}\in\mathbb R^d$, PR computes coordinate-wise moments over the patient's available channels:

\begin{equation}
\begin{aligned}\mu_{pj}&=\frac1{C_p}\sum_{c\in\mathcal C_p}z_{pcj},\\\sigma_{pj}^2&=\frac1{C_p}\sum_{c\in\mathcal C_p}(z_{pcj}-\mu_{pj})^2,\end{aligned}
\label{eq:5}
\end{equation}

\begin{equation}
\mathcal R_p(z_{pcj})=\frac{z_{pcj}-\mu_{pj}}{\sigma_{pj}}.
\label{eq:6}
\end{equation}
Equation (6) applies to supported, nonconstant coordinates. It specifies where a channel lies in the patient's observed distribution using the observable channel set without its clinical labels. For a linear predictor, its effect is explicit. Writing $D_p=\operatorname{diag}(\sigma_p)$ on supported coordinates,

\begin{equation}
w^{\mathsf T}\mathcal R_p(z)+b=(D_p^{-1}w)^{\mathsf T}z+b-w^{\mathsf T}D_p^{-1}\mu_p.
\label{eq:7}
\end{equation}
A shared predictor therefore acquires patient-dependent coordinate weights and an offset. In multiple dimensions, the changed coordinate weights can alter within-patient channel order; the operator is not limited to shifting a decision threshold. For the transformation in Eq. (6), positive affine recoding of a supported coordinate leaves its standardized value unchanged.

For physiological features, a coordinate is supported when its patient SD exceeds $10^{-8}$ times its FIT-derived population scale; otherwise it is assigned zero. Moments use available finite values. After patient standardization, missing values are imputed and coordinates are standardized using an imputer and scaler fitted on FIT patients only. Thus the order is patient-relative transformation followed by FIT-derived preprocessing, and PR replaces the original 12 coordinates rather than doubling their dimension. Raw-representation PR computes moments in double precision and replaces a denominator at or below $10^{-8}$ with 1; constant coordinates map to zero.

To test reference transfer independently of classification, each patient undergoes 100 fixed, label-independent half-channel partitions. One half supplies the patient moments and the other is assessed, then the roles are exchanged. The comparison uses identical partitions and training-standardized evaluation coordinates. Global and same-source references use training-channel moments. Location-only uses the reference-half patient mean with global SD; scale-only uses global mean with reference-half patient SD; full PR uses both patient moments. Each defines a marginal Gaussian forecast for an unchanged held-out value x, scored by the continuous ranked probability score (CRPS) \cite{ref35}:

\begin{equation}
\operatorname{CRPS}(\mu,\sigma;x)=\sigma\left[u(2\Phi(u)-1)+2\phi(u)-\pi^{-1/2}\right],
\label{eq:8}
\end{equation}
where $u=(x-\mu)/\sigma$ and $\Phi$, $\phi$ are the standard normal distribution and density. At $\sigma=0$, CRPS equals $|x-\mu|$. We average coordinates, held-out channels, and repetitions within patients before averaging patients equally. Each of the 80 patients is assessed once under the reference experiment's fixed allocation; global and source moments use its corresponding training patients. Source-stratified paired patient bootstrap with 10,000 resamples provides nominal 95\% intervals.

\subsection{DREChanNet for Ictal Localization}
The physiological branch applies feature PR to $f_{pc}$. The raw branch encodes valid windows with a shared two-layer temporal CNN: 12 and 24 output channels, kernel sizes 9 and 7, stride 4, GELU after each convolution, and global temporal average pooling. Before encoding, each waveform is centered, divided by its own population SD with a $10^{-6}$ floor, and clipped to [-8, 8]. The resulting embedding $h_{pcsw}$ has 24 coordinates.

For each channel and seizure, we subtract its mean pre-onset embedding:

\begin{equation}
\Delta h_{pcsw}=h_{pcsw}-\frac1{|\mathcal B_{pcs}|}\sum_{u\in\mathcal B_{pcs}}h_{pcsu}.
\label{eq:9}
\end{equation}
Separate averages over the onset-relative stages $I_1=[0,8)$ s and $I_2=[8,20)$ s retain early and later ictal activity:

\begin{equation}
a_{pcsk}=\frac{\sum_w m_{pcsw}\mathbf1[t_w\in I_k]\Delta h_{pcsw}}{\sum_w m_{pcsw}\mathbf1[t_w\in I_k]},\quad k\in\{1,2\},
\label{eq:10}
\end{equation}
where m is the validity mask and the expression is defined for nonempty stages. Across available seizures, each stage contributes its mean and population SD, producing a 2 $\times$ 2 $\times$ 24 = 96-dimensional raw vector $r_{pc}$. Empty stages do not contribute to pooling. Raw PR transforms this vector across the patient's channels.

Let $\widetilde r_{pc}\in\mathbb R^{96}$ denote the processed raw vector and $\widetilde f_{pc}\in\mathbb R^{14}$ the processed physiological vector, whose 12 active coordinates occupy a fixed-width implementation with two zero coordinates. A shared 64-unit GELU head produces the NEZ logit:

\begin{equation}
\begin{aligned}h_{pc}^{\mathrm{joint}}&=\operatorname{GELU}(\alpha_rW_r\widetilde r_{pc}+\alpha_fW_f\widetilde f_{pc}+b),\\\ell_{pc}&=w^{\mathsf T}h_{pc}^{\mathrm{joint}}+b_o,\quad q_{pc}=\operatorname{sigmoid}(-\ell_{pc}),\end{aligned}
\label{eq:11}
\end{equation}
with fixed dimension-based scales $\alpha_r=\sqrt{110/192}$ and $\alpha_f=\sqrt{110/28}$. The two branches therefore contribute additively before the shared nonlinearity. DREChanNet contains 9,329 trainable parameters. Branch-only variants retain the corresponding input block and head; component ablations remove feature PR, raw PR, both PR operations, or the raw pre-onset reference; the physiological DELTA construction is retained in that ablation.

\subsection{Learning and Integration With Existing Models}
Training assigns equal weight to patients using one complete patient per optimizer update. For $\rho=N_{\mathrm{EZ}}^{\mathrm{FIT}}/N_{\mathrm{NEZ}}^{\mathrm{FIT}}$, the patient loss is

\begin{equation}
\mathcal L_p^{\mathrm{BCE}}=-\frac1{C_p}\sum_c[y_{pc}\log q_{pc}+\rho(1-y_{pc})\log(1-q_{pc})].
\label{eq:12}
\end{equation}
For class $k\in\{\mathrm{EZ},\mathrm{NEZ}\}$, let $t_{pc}^k$ be its binary target and $s_{pc}^k$ its predicted probability: $(t^{\mathrm{EZ}},s^{\mathrm{EZ}})=(y,q)$ and $(t^{\mathrm{NEZ}},s^{\mathrm{NEZ}})=(1-y,1-q)$. Its soft F1 is

\begin{equation}
F_{p,k}^{\mathrm{soft}}=\frac{2\sum_c t_{pc}^ks_{pc}^k}{\max(\sum_c t_{pc}^k+\sum_c s_{pc}^k,10^{-8})}.
\label{eq:13}
\end{equation}
The combined objective is

\begin{equation}
\mathcal L_p=0.75\mathcal L_p^{\mathrm{BCE}}+0.25\left[1-\frac{F_{p,\mathrm{EZ}}^{\mathrm{soft}}+F_{p,\mathrm{NEZ}}^{\mathrm{soft}}}{2}\right].
\label{eq:14}
\end{equation}
The mean of $\mathcal L_p$ across FIT patients is the training objective. AdamW uses learning rate 3 $\times$ $10^{-4}$ and weight decay $10^{-3}$, with gradient clipping at 1. Training has a 30-epoch budget, a six-epoch minimum, and six-epoch stopping patience. Validation selects the checkpoint and a threshold on the NEZ-probability grid 0, 0.005, . . . , 1, prioritizing patient-mean macro-F1 and then EZ-F1. A channel is called EZ when 1-$q_{pc}$ is below that threshold.
Feature baselines comprise linear SVM, random forest, XGBoost, and an MLP. They use the same 12 coordinates, replaced by their patient-standardized values in the PR condition. Required imputers and population scalers are fitted on FIT patients. Raw baselines are channel-level adaptations of TimeConv-CNN \cite{ref24}, EEGNet \cite{ref36}, SEEG-Net \cite{ref15}, and SEEGformer \cite{ref29} to the common 2-s window protocol. Their channel embeddings are formed by averaging windows within seizures and retaining cross-seizure mean and SD. Their frozen original EZ logit $a_{pc}^0$ receives a learned linear PR correction:

\begin{equation}
a_{pc}^{\mathrm{PR}}=a_{pc}^0+\beta^{\mathsf T}\mathcal R_p(h_{pc})+b.
\label{eq:15}
\end{equation}
Zero initialization recovers the original predictor. Only the correction and its validation threshold are fitted; the encoder and original classifier remain fixed. The correction uses patient-equal weighted BCE and AdamW with learning rate and weight decay $10^{-3}$. The TimeConv seizure-sufficiency analysis retains its original end-to-end PR readout, as identified in its table.

\subsection{PR Adaptation to Interictal Recordings}

\begin{table}[!htb]
\centering\caption{Patient-based interictal subset of Omni-iEEG after channel filtering and deduplication.}\label{tab:omni_cohort}
\footnotesize\setlength{\tabcolsep}{3pt}
\begin{tabular}{@{}lp{.50\columnwidth}@{}}\toprule
Characteristic & Evaluation cohort\\\midrule
Patients & 124\\
Interictal recordings & 137 EDF files\\
Unique patient--channel pairs & 10,202\\
Pathological / normal & 1,113 / 9,089\\
Sources (patients) & Open-iEEG (103), HUP (10), SourceSink (11)\\
Evaluation & Five patient-held-out folds\\\bottomrule
\end{tabular}\end{table}

We construct a patient-based evaluation from Omni-iEEG \cite{ref24}, comprising 124 patients, 137 interictal records, and 10,202 labeled canonical patient--channel pairs (Table 2). Repeated observations of a patient's channel are combined into one evaluation unit. Good, nonresected channels from seizure-free patients are labeled normal; otherwise, good SOZ channels are labeled pathological. Remaining good channels provide unlabeled reference context. Patients with both labeled classes are included. These eligibility rules and channel consolidation define a different evaluation set from the published Task 2 cohort. The original benchmark uses a subject-level 60\%/40\% split; we use five patient-held-out folds with equal-patient endpoints.

Each 60-s recording at 1 kHz is divided into 30 nonoverlapping 2-s segments. The frozen TimeConv backbone used for this evaluation yields 32-dimensional embeddings and logits. Repeated records are averaged within each segment, followed by averaging across segments. PR uses all observable good channels, including unlabeled context, with denominator $\sqrt{\operatorname{Var}(h_p)+10^{-5}}$. The pathological-channel logit is

\begin{equation}
a_{pc}^{\mathrm{PR}}=a_{pc}^0+\beta^{\mathsf T}\widetilde h_{pc}+b_{g(p)},
\label{eq:16}
\end{equation}
where g(p) identifies one of three sources. The 32 coefficients and three source biases add 35 zero-initialized parameters. Patient-wise AdamW uses learning rate $10^{-3}$, weight decay $10^{-4}$, and pathological-class weight 2, with an additional $10^{-3}$ quadratic penalty on the coefficients and source-frequency-weighted biases.

Within each outer FIT population, threefold patient-level cross-validation selects the readout epoch from 50 epochs using AUROC and then AP. The selected epoch's inner out-of-fold readout predictions determine its classification threshold, after which the readout is fitted on all outer FIT patients. This inner validation concerns the readout; the frozen backbone was trained on outer FIT patients. The baseline threshold uses its outer FIT predictions. Outer patients are excluded from backbone and readout fitting.

\subsection{Evaluation Measures and Additional Analyses}
Classification and ranking metrics are calculated within each patient and then averaged equally across patients. Repeated-run summaries first average each patient's endpoint across runs. We report accuracy, AUROC, average precision, macro-F1, class-specific F1, mean reciprocal rank (MRR), and Top-1 hit rate. Ranking follows decreasing EZ scores. If $E_p(K)$ is the number of annotated EZ channels among the first K channels, precision is $E_p(K)$/K, recall is $E_p(K)$/$n_{p,\mathrm{EZ}}$, and hit rate is $\mathbf{1}[E_p(K)>0]$. Reciprocal rank is the inverse rank of the first annotated EZ channel.

Source-held-out evaluation excludes an entire source from training and reports the displayed four-run descriptive average, the equal-source mean, and the lowest source mean. Seizure sufficiency uses the same 57 patients with at least three seizures and 5,735 common channels, comparing one, two, and all available seizures. All seizure subsets are evaluated when at most 50 exist; otherwise 50 fixed subsets are sampled. Checkpoints and thresholds remain frozen, and subset endpoints are averaged within patients. The Top-K figure uses one frozen run per model and nominal 95\% source-stratified patient-bootstrap intervals.

\subsection{Statistical Analysis}
The analysis uses patient-equal summaries. The ictal cohort includes 80 patients and the interictal cohort includes 124 patients; the seizure-sufficiency analysis includes 57 patients. Feature association uses two-sided Wilcoxon signed-rank tests with Benjamini--Hochberg correction across 12 coordinates. Nominal 95\% patient-bootstrap intervals are used for feature associations, and source-stratified paired patient-bootstrap intervals are used for the reference comparison; each uses 10,000 resamples. Data preprocessing, metrics, and additional resampling procedures are specified above.

\section{Results}

\begin{figure*}[!tb]
\centering
\includegraphics[width=\linewidth]{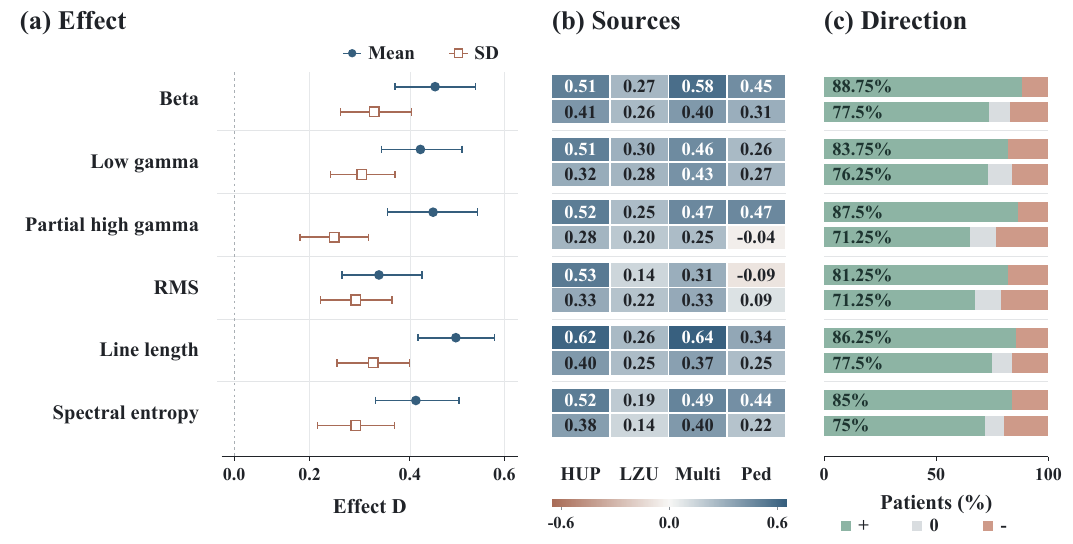}
\caption{Within-patient associations between physiological descriptors and EZ annotations. (a) Patient-mean signed rank-biserial effects, D, with nominal 95\% patient-bootstrap confidence intervals. Paired symbols denote cross-seizure means and standard deviations (SDs). (b) Mean effects in each data source. (c) Fractions of patients with positive, zero, or negative effects; labels give the positive fraction. All 80 patients are included.}
\label{fig:3}
\end{figure*}

\begin{table}[!htb]
\centering
\caption{Prediction of unseen channels from label-free reference channels. CRPS is evaluated in shared training-standardized coordinates; intervals are nominal 95\% patient-bootstrap confidence intervals.}
\label{tab:crps_results}
\begin{tabular}{lcc}
\toprule
Method & CRPS $\downarrow$ & 95\% CI \\
\midrule
Global Reference     & 0.4509 & [0.3793, 0.5373] \\
Same-Source Reference & 0.4526 & [0.3783, 0.5414] \\
Location-only           & 0.3132 & [0.2858, 0.3480] \\
Scale-only            & 0.4357 & [0.3507, 0.5372] \\
\textbf{Full PR}      & \textbf{0.1734} &
\textbf{[0.1315, 0.2267]} \\
\bottomrule
\end{tabular}
\end{table}


\definecolor{baselinebg}{RGB}{234,241,248}
\definecolor{pluginbg}{RGB}{232,245,236}
\definecolor{ourbg}{RGB}{249,237,231}
\definecolor{improvebg}{RGB}{249,246,234}
\definecolor{bestred}{RGB}{172,25,39}

\begin{table*}[!tb]
\centering
\caption{Comparison on the 80-patient localization cohort using five patient-held-out outer folds. Results are patient-equal means (\%). }
\label{tab:task1_main}
\setlength{\tabcolsep}{4pt}
\renewcommand{\arraystretch}{1.10}
\scriptsize
\resizebox{\textwidth}{!}{%
\begin{tabular}{ll*{8}{r}}
\toprule
\textbf{Methods} & \textbf{Input} &
\textbf{ACC}$\uparrow$ & \textbf{AUROC}$\uparrow$ & \textbf{AP}$\uparrow$ &
\textbf{Macro-F1}$\uparrow$ & \textbf{EZ-F1}$\uparrow$ & \textbf{NEZ-F1}$\uparrow$ &
\textbf{MRR}$\uparrow$ & \textbf{Top-1}$\uparrow$ \\
\midrule
\rowcolor{baselinebg}
\multicolumn{10}{c}{\textbf{Original Baselines}} \\
Linear SVM & 12D & 64.19 & \textbf{68.65} & \textbf{48.16} & 46.06 & 28.73 & 63.40 & 64.83 & 53.75 \\
Random Forest & 12D & 62.95 & 60.95 & 42.76 & 48.71 & \textbf{31.32} & 66.09 & 61.74 & 50.00 \\
XGBoost & 12D & 67.67 & 63.02 & 44.91 & \textbf{51.71} & 30.79 & 72.63 & 64.05 & 52.50 \\
MLP & 12D & 65.65 & 66.11 & 47.63 & 49.67 & 30.01 & 69.32 & \textbf{68.39} & \textbf{58.75} \\
TimeConv-CNN & Raw & 66.76 & 62.06 & 43.15 & 49.11 & 26.70 & 71.53 & 57.01 & 43.75 \\
EEGNet & Raw & 67.94 & 56.01 & 34.52 & 44.17 & 12.57 & 75.77 & 47.50 & 30.00 \\
SEEG-Net & Raw & \textbf{72.85} & 45.06 & 26.38 & 42.22 & 3.08 & \textbf{81.37} & 34.10 & 22.50 \\
SEEGformer & Raw & 62.59 & 53.45 & 31.88 & 42.45 & 15.96 & 68.95 & 44.38 & 26.25 \\
\midrule
\rowcolor{pluginbg}
\multicolumn{10}{c}{\textbf{Baselines + Patient-Relative Methods}} \\
Linear SVM + PR & 12D & 72.28 & \textbf{74.36} & \textbf{54.32} & \textbf{62.47} & \textbf{45.21} & 79.72 & \textbf{73.88} & \textbf{65.00} \\
Random Forest + PR & 12D & \textbf{74.90} & 71.14 & 51.00 & 59.72 & 37.46 & \textbf{81.99} & 68.29 & 60.00 \\
XGBoost + PR & 12D & 74.45 & 71.42 & 51.54 & 59.26 & 37.02 & 81.49 & 73.44 & 63.75 \\
MLP + PR & 12D & 73.13 & 71.53 & 51.19 & 60.21 & 39.86 & 80.56 & 70.56 & 57.50 \\
TimeConv-CNN + PR & Raw & 67.47 & 64.68 & 44.58 & 54.87 & 35.91 & 73.83 & 60.69 & 45.00 \\
EEGNet + PR & Raw & 69.38 & 58.93 & 35.99 & 50.10 & 20.66 & 79.53 & 52.46 & 38.75 \\
SEEG-Net + PR & Raw & 67.68 & 60.18 & 37.12 & 51.29 & 24.21 & 78.36 & 54.15 & 38.75 \\
SEEGformer + PR & Raw & 64.19 & 55.76 & 33.98 & 48.46 & 22.72 & 74.20 & 50.30 & 32.50 \\
\rowcolor{improvebg}
\multicolumn{2}{l}{\textbf{Average Improvement}} & +4.11 & +6.58 & +5.04 & +9.03 & +10.49 & +7.58 & +7.72 & +7.97 \\
\midrule
\rowcolor{ourbg}
\multicolumn{10}{c}{\textbf{Our Method}} \\
DREChanNet-12D & 12D & 75.08 & 74.96 & 54.20 & 62.49 & 42.46 & 82.51 & 74.49 & 64.58 \\
DREChanNet-Raw & Raw & 75.70 & 72.88 & 53.59 & 61.47 & 39.26 & \textcolor{bestred}{\textbf{83.68}} & 76.23 & 67.92 \\
\textbf{DREChanNet} & Full & \textcolor{bestred}{\textbf{76.05}} & \textcolor{bestred}{\textbf{76.67}} & \textcolor{bestred}{\textbf{57.08}} & \textcolor{bestred}{\textbf{64.38}} & \textcolor{bestred}{\textbf{45.16}} & 83.59 & \textcolor{bestred}{\textbf{78.77}} & \textcolor{bestred}{\textbf{69.58}} \\
\midrule
\rowcolor{improvebg}
\multicolumn{2}{l}{\textbf{Improvement vs Baseline}} & +3.20 & +8.01 & +8.92 & +12.67 & +13.84 & +2.22 & +10.38 & +10.83 \\
\bottomrule
\end{tabular}%
}

\end{table*}

\subsection{Physiological Evidence Within Patients}
The six descriptors provide a physiological basis for channel comparison (Fig. 3). All 12 coordinates have positive overall signed rank-biserial effects, and the mean coordinate of each descriptor has a larger effect than its corresponding SD. Positive effects occur in 81.25--88.75\% of patients for the mean coordinates and 71.25--77.50\% for the SD coordinates. Beta-power mean is positively associated with EZ labels in 88.75\% of patients, followed by partial high-gamma-power mean in 87.50\% and line-length mean in 86.25\%. The source-wise panel shows how the magnitudes vary across cohorts. Together, the panels motivate comparing physiological summaries within each patient's implanted set.

\subsection{Predictive Value of the Patient Reference}
Patient-specific reference statistics predict the distribution of held-out channels more accurately than the population references in Table 3. Full PR achieves a CRPS of 0.1734, compared with 0.4509 for the global reference, a 61.5\% reduction. The Location-only and Scale-only references yield 0.3132 and 0.4357, respectively; using both patient statistics gives the lowest error. The same-source reference scores 0.4526. Because the channels used to estimate each patient reference are separated from the channels being scored, this result supports the reference's predictive value beyond the channels from which it was estimated.

\subsection{PR Across Baselines and DREChanNet}
Adding PR increases macro-F1 in all eight baseline comparisons in Table 4. The mean gains across these paired comparisons are 9.03 percentage points in macro-F1, 10.49 points in EZ-F1, and 6.58 points in AUROC. Linear SVM improves from 46.06\% to 62.47\% macro-F1, while the frozen TimeConv-CNN with a PR residual improves from 49.11\% to 54.87\%. These comparisons demonstrate the utility of patient-relative inputs in feature classifiers and patient-relative corrections in learned representations.

DREChanNet combines these two forms of evidence. It reaches 64.38\% macro-F1, 76.67\% AUROC, 57.08\% average precision, 78.77\% MRR, and 69.58\% Top-1 accuracy. Its macro-F1 is 12.67 percentage points above the strongest original baseline and 1.91 points above the strongest PR-augmented baseline. The feature-only and raw-only variants reach 62.49\% and 61.47\% macro-F1, respectively, compared with 64.38\% for their joint model.

\subsection{Contributions of the Model Components}

\begin{table}[!htb]
\centering
\caption{Component ablation of DREChanNet.}
\label{tab:drechannet_ablation}
\footnotesize
\setlength{\tabcolsep}{3.1pt}
\renewcommand{\arraystretch}{1.12}
\begin{tabular}{@{}lrrrrr@{}}
\toprule
\textbf{Configuration} & \textbf{F1} & \textbf{AUC} & \textbf{AP} & \textbf{EZ-F1} & \textbf{F1} \\
\midrule
\rowcolor{black!5} \textbf{Full} & \textbf{64.38} & \textbf{76.67} & \textbf{57.08} & \textbf{45.16} & \textbf{69.58} \\
w/o Feature PR & 62.15 & 73.65 & 54.38 & 40.71 & 68.75 \\
w/o Raw PR & 62.72 & 76.36 & 56.17 & 43.90 & 66.25 \\
w/o Both PR & 49.26 & 74.97 & 55.41 & 37.08 & 67.08 \\
w/o Pre-onset & 61.70 & 73.17 & 53.58 & 41.03 & 67.92 \\
Raw-only & 61.47 & 72.88 & 53.59 & 39.26 & 67.92 \\
Feature-only & 62.49 & 74.96 & 54.20 & 42.46 & 64.58 \\
\bottomrule
\end{tabular}
\end{table}

The ablation results connect the joint model's performance to its reference operations (Table 5). Removing feature PR changes macro-F1 from 64.38\% to 62.15\%, and removing raw PR changes it to 62.72\%. Removing both gives 49.26\%, corresponding to a 15.11-percentage-point difference computed from the unrounded scores. Each branch alone also yields lower macro-F1 than the joint model. The raw branch's same-channel pre-onset contrast contributes a further distinction: without it, macro-F1 is 61.70\% and AUROC is 73.17\%, compared with 64.38\% and 76.67\% for the full model. The component comparisons therefore support combining physiological and learned temporal representations with both temporal and patient-level references.

\subsection{Source Transfer and Seizure Availability}

\begin{table*}[!tb]
\centering
\caption{Exploratory leave-one-source-out localization across four sources, summarized over four reported repetitions. Center mean weights the sources equally; worst is the lowest source mean. Scores are proportions.}
\label{tab:loco_generalization}

\setlength{\tabcolsep}{7pt}
\renewcommand{\arraystretch}{1.18}
\footnotesize

\begin{tabular}{llcccccc}
\toprule
& & \multicolumn{4}{c}{\textbf{Held-out Source}}
& \multicolumn{2}{c}{\textbf{Summary}} \\
\cmidrule(lr){3-6}
\cmidrule(lr){7-8}

\textbf{Metric} &
\textbf{Method} &
\textbf{HUP} &
\textbf{LZU} &
\textbf{Multicenter} &
\textbf{Pediatric} &
\textbf{Center Mean} &
\textbf{Worst} \\
\midrule

\multirow{5}{*}{\textbf{Macro-F1}}
& TimeConv-CNN
& 0.4066 & 0.4701 & 0.5000 & 0.3217
& 0.4246 & 0.3217 \\

& SVM + PR
& \textbf{0.6623} & 0.5240 & 0.6148 & 0.5950
& 0.5990 & 0.5240 \\

\addlinespace[2pt]
\cmidrule(lr){2-8}
\addlinespace[2pt]

& Raw-only
& 0.6098 & 0.5376 & 0.6316 & 0.6111
& 0.5975 & 0.5376 \\

& Feature-only
& 0.6583 & 0.5633 & 0.5988 & 0.5862
& 0.6017 & 0.5633 \\

\rowcolor{black!6}
& \textbf{DREChanNet (Full)}
& 0.6498 & \textbf{0.5649}
& \textbf{0.6482} & \textbf{0.6167}
& \textbf{0.6199} & \textbf{0.5649} \\

\midrule

\multirow{5}{*}{\textbf{AUROC}}
& TimeConv-CNN
& 0.5341 & 0.5752 & 0.7235 & 0.5216
& 0.5886 & 0.5216 \\

& SVM + PR
& 0.7866 & 0.6266 & \textbf{0.8228} & 0.6575
& 0.7234 & 0.6266 \\

\addlinespace[2pt]
\cmidrule(lr){2-8}
\addlinespace[2pt]

& Raw-only
& 0.7434 & 0.6273 & 0.7607 & 0.6951
& 0.7066 & 0.6273 \\

& Feature-only
& \textbf{0.7946} & 0.6482 & 0.8220 & 0.6727
& 0.7344 & 0.6482 \\

\rowcolor{black!6}
& \textbf{DREChanNet (Full)}
& 0.7813 & \textbf{0.6652}
& 0.8071 & \textbf{0.7062}
& \textbf{0.7400} & \textbf{0.6652} \\

\bottomrule
\end{tabular}

\end{table*}

In the source-held-out evaluation, DREChanNet attains the highest equal-source mean and worst-source value for both reported metrics (Table 6). The corresponding macro-F1 values are 0.6199 and 0.5649, and the AUROC values are 0.7400 and 0.6652. Full-model macro-F1 ranges from 0.5649 to 0.6498 across the four held-out sources. The joint model also exceeds either individual branch in the equal-source summary. These descriptive results assess transfer across HUP, LZU, the public multicenter source, and the pediatric Fudan source.

\begin{table}[!htb]
    \centering
    \caption{Seizure sufficiency using frozen models on the same 57 patients and 5,735 common channels. Scores are percentages. TimeConv uses its original end-to-end PR readout in this analysis.}
\label{tab:seizure_sufficiency}
    \footnotesize
    \setlength{\tabcolsep}{3pt}
    \renewcommand{\arraystretch}{1.08}
    \begin{tabular}{@{}lccrrr@{}}
        \toprule
        Model & Seizures & PR & Macro-F1 & AUROC & EZ-F1 \\
        \midrule
        \multirow{6}{*}{SVM}
          & \multirow{2}{*}{1} & No  & 47.29 & 69.11 & 19.18 \\
          &                   & Yes & 60.64 & 71.90 & 41.36 \\
        \addlinespace[1pt]
          & \multirow{2}{*}{2} & No  & 47.63 & 68.69 & 26.44 \\
          &                   & Yes & 61.42 & 73.43 & 43.51 \\
        \addlinespace[1pt]
          & \multirow{2}{*}{All} & No  & 46.13 & 69.79 & 27.48 \\
          &                     & Yes & 62.08 & 74.74 & 45.32 \\
        \midrule
        \multirow{6}{*}{TimeConv}
          & \multirow{2}{*}{1} & No  & 48.35 & 60.81 & 19.86 \\
          &                   & Yes & 49.18 & 63.15 & 17.37 \\
        \addlinespace[1pt]
          & \multirow{2}{*}{2} & No  & 49.49 & 60.81 & 27.18 \\
          &                   & Yes & 49.95 & 58.92 & 22.12 \\
        \addlinespace[1pt]
          & \multirow{2}{*}{All} & No  & 50.27 & 61.67 & 30.56 \\
          &                     & Yes & 49.85 & 58.86 & 21.54 \\
        \midrule
        \multirow{6}{*}{DREChanNet}
          & \multirow{2}{*}{1} & No  & 50.06 & 71.02 & 27.26 \\
          &                   & Yes & 60.94 & 71.50 & 38.62 \\
        \addlinespace[1pt]
          & \multirow{2}{*}{2} & No  & 50.40 & 74.95 & 34.04 \\
          &                   & Yes & 62.26 & 74.08 & 41.41 \\
        \addlinespace[1pt]
          & \multirow{2}{*}{All} & No  & 49.54 & 76.55 & 35.76 \\
          &                     & Yes & 63.34 & 75.65 & 43.48 \\
        \bottomrule
    \end{tabular}
\end{table}

The seizure-sufficiency analysis holds the evaluated patients, channels, checkpoints, and thresholds constant while varying the number of seizures available at inference (Table 7). For DREChanNet with PR, macro-F1 increases from 60.94\% with one seizure to 62.26\% with two and 63.34\% with all available seizures. EZ-F1 increases from 38.62\% to 41.41\% and 43.48\%, respectively. The corresponding macro-F1 values without PR are 50.06\%, 50.40\%, and 49.54\%. Thus, a single seizure already supports localization, while additional seizures improve the reported model's balanced classification and recovery of annotated EZ channels.

\subsection{Channel Prioritization and an Illustrative PR Case}

\begin{figure}[!htb]
\centering
\includegraphics[width=\linewidth]{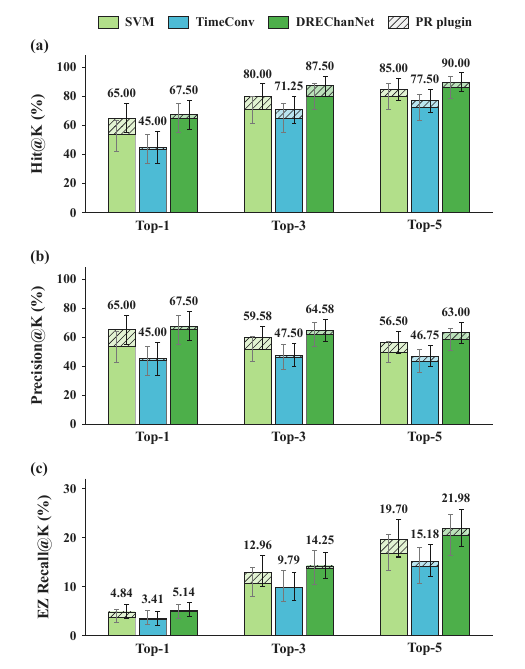}
\caption{Channel prioritization for SVM, TimeConv-CNN, and DREChanNet. (a) Hit rate, (b) precision, and (c) EZ recall at the first one, three, and five channels. Solid bars show the baseline and hatched extensions the PR increment; labels give the PR result. Error bars are nominal 95\% source-stratified patient-bootstrap intervals. Each model uses one frozen run, whereas Table 4 averages the reported DREChanNet runs.}
\label{fig:4}
\end{figure}

\begin{figure*}[!tb]
\centering
\includegraphics[width=\linewidth]{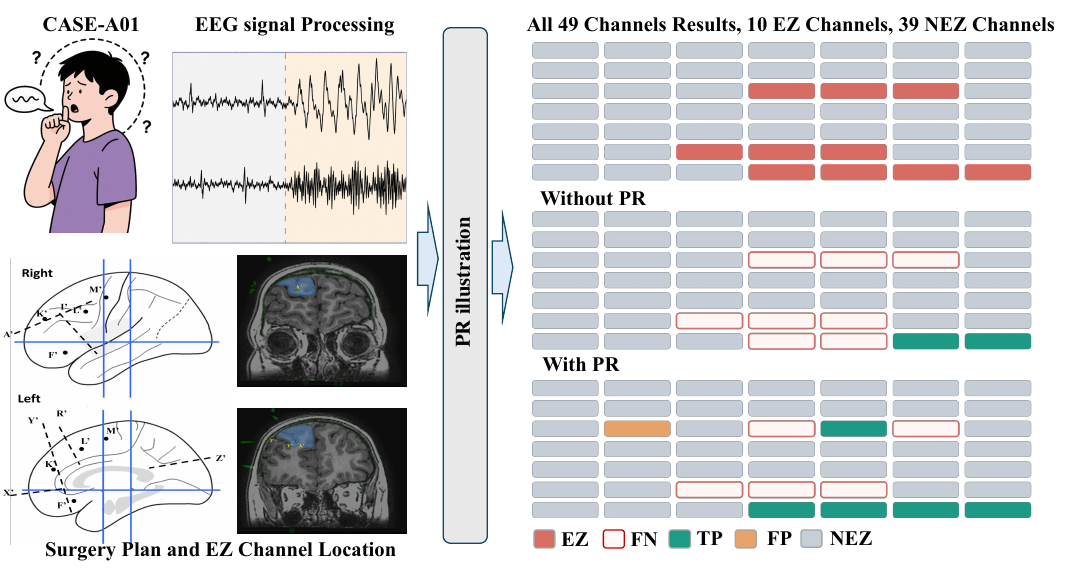}
\caption{Illustrative PR comparison for CASE-A01: 49 channels, including 10 annotated EZ and 39 NEZ channels. Recording and surgical-planning panels illustrate the workflow.}
\label{fig:5}
\end{figure*}
In the frozen-run ranking analysis, DREChanNet with PR places at least one EZ channel among the first one, three, and five channels in 67.50\%, 87.50\%, and 90.00\% of patients, respectively (Fig. 4). At five reviewed channels, patient-mean precision is 63.00\% and EZ recall is 21.98\%. The corresponding PR results are 56.50\% precision and 19.70\% recall for SVM, and 46.75\% and 15.18\% for TimeConv-CNN. These measures quantify the concentration of annotated EZ channels in a short review list.

Figure 5 provides a channel-level illustration of the PR comparison using the separately identified development encoder. In this case, the number of correctly detected EZ channels increases from 2 of 10 to 5 of 10, with one NEZ channel assigned a positive prediction after PR. The fixed channel order makes the changed decisions directly visible alongside the clinical annotations.

\subsection{Application to Interictal Recordings}

\begin{table}[!htb]
\centering
\caption{Patient-equal interictal performance on Omni-iEEG under patient-held-out fivefold evaluation.}
\label{tab:omni_pr}
\footnotesize
\setlength{\tabcolsep}{4pt}
\renewcommand{\arraystretch}{1.08}
\begin{tabular}{lccc}
\toprule
Metric & Baseline & +PR & Improvement \\
\midrule
AUROC $\uparrow$ & 0.7622 & \textbf{0.8118} & \textbf{+4.97 pp} \\
AP $\uparrow$ & 0.4773 & \textbf{0.5235} & \textbf{+4.63 pp} \\
Macro-F1 $\uparrow$ & 0.6104 & \textbf{0.6558} & \textbf{+4.55 pp} \\
Pathological-F1 $\uparrow$ & 0.3135 & \textbf{0.3923} & \textbf{+7.88 pp} \\
MRR $\uparrow$ & 0.6075 & \textbf{0.6590} & \textbf{+5.15 pp} \\
Top-1 $\uparrow$ & 0.5081 & \textbf{0.5645} & \textbf{+5.65 pp} \\
\bottomrule
\end{tabular}
\end{table}

The patient-based Omni-iEEG evaluation uses the cohort specified in Table 2: 124 patients, 137 recordings, and 10,202 unique labeled patient--channel pairs. With a frozen TimeConv backbone, adding the PR readout improves all six reported endpoints (Table 8). AUROC increases from 0.7622 to 0.8118, average precision from 0.4773 to 0.5235, and macro-F1 from 0.6104 to 0.6558. Pathological-channel F1 increases from 0.3135 to 0.3923; MRR increases from 0.6075 to 0.6590, and Top-1 accuracy from 0.5081 to 0.5645. These results extend the evaluated application of the PR plug-in to interictal representations, without requiring the ictal architecture or its pre-onset input.

\section{Discussion}
Computational localization faces a coupled problem: the clinical target is inferred, the brain is sampled selectively, and the meaning of a measured channel depends on the patient in whom it was recorded. Recent reviews identify these relationships among the main obstacles to translating quantitative iEEG into surgical practice \cite{ref17,ref18,ref19}. Our study connects them through a defined sequence of evidence: physiological association, prediction of a patient's channel distribution, PR-based localization, and evaluation across patients, recording conditions, and seizure states.

\subsection{Clinical Targets and the Meaning of Ground Truth}

\begin{table}[!htb]\centering
\caption{Clinical observations used to assess localization and their interpretation.}\label{tab:clinical_targets}
\footnotesize\setlength{\tabcolsep}{3pt}\renewcommand{\arraystretch}{1.15}
\begin{tabular}{@{}p{.23\columnwidth}p{.71\columnwidth}@{}}\toprule
Observation & Meaning for validation\\\midrule
Clinical SOZ & Interpreted onset within sampled contacts; its definition may include early propagation \cite{ref9}.\\
Resection or ablation & The intervention actually delivered; agreement measures overlap with treatment, with contact-specific necessity still undetermined \cite{ref38}.\\
Seizure outcome & Patient-level response at a defined follow-up; supports confidence in localization without specifying a unique channel map \cite{ref10}.\\
Algorithmic abnormality & A signal-derived deviation or score; its interpretation depends on the physiological measurement and reference distribution \cite{ref17,ref21}.\\\bottomrule
\end{tabular}\end{table}

The EZ is inferred from several observations that describe different biological and clinical objects. Its surgical definition concerns tissue whose treatment is necessary for seizure freedom, whereas the SEEG tradition also emphasizes the spatial and temporal organization of the electroclinical seizure \cite{ref5,ref37}, \cite{ref38}. These distinctions remain consequential in current practice. In a survey of 321 experts from 39 countries, 55.8\% defined the SOZ by onset alone, while 42.1\% included early propagation \cite{ref9}. A binary channel label can therefore encode different clinical definitions even when datasets use the same name.

Resection introduces another distinction. Seizure freedom after treating a region supports the adequacy of that intervention at the patient level. It does not identify which individual contacts within the treated region were necessary for the result. Anatomical constraints and the planned procedure can produce a treatment region extending beyond the tissue essential to seizure generation \cite{ref7,ref38}. Assigning every resected channel an EZ label can consequently enlarge the positive set. Outcome supplies an additional validation dimension: a retrospective study found that the fraction of the observed SOZ resected did not closely track surgical outcome \cite{ref39}. The treatment map, observed seizure onset, and outcome should therefore remain distinguishable quantities (Table 9).

This is structured uncertainty arising from the clinical definition, implantation and intervention. An algorithm that concentrates scores on a smaller subset can lose recall against a broad resection label, while an algorithm that follows that label closely can learn aspects of local treatment practice. Retaining the original observations makes such disagreement available for clinical adjudication. The issue also begins before annotation: human microelectrode recordings distinguish intensely recruited seizure cores from surrounding tissue displaying large electrographic signals \cite{ref8}, and a systematic review found limited evidence establishing the reliability of visual SOZ delineation \cite{ref40}. A contact's recorded signal, assigned label and role in seizure generation are related stages of inference.

An actionable response is to preserve SOZ annotations, resection or ablation maps, electrode coordinates, follow-up outcomes and localization confidence as separate data fields. The Delphi framework of Ryvlin et al. provides a structured way to describe confidence in reported EZ localization \cite{ref10}. These fields permit analyses of where a model agrees with onset interpretation, where it agrees with treatment, and whether either agreement relates to outcome. They also make disagreements available for adjudication instead of collapsing all clinical evidence into an apparently exact channel target.

\subsection{Making Patient-Level Evidence Comparable}
The target definition and the evaluation unit jointly determine what a reported score means. Windows from the same channel, seizures from the same implantation, and channels from the same patient share substantial context. A split across those observations estimates a different use case from prediction for an unseen patient \cite{ref32}. Weighting also changes the clinical question. Pooled channel metrics emphasize patients with more contacts; equal-patient metrics summarize performance for a typical included patient. These choices can change method rankings even when the underlying predictions remain fixed.

This distinction is particularly relevant when combining clinical sources. Quantitative iEEG reviews describe sensitivity to pathology, electrode placement and dataset construction \cite{ref17,ref19}. A 2026 ILAE meta-analysis of connectivity-based localization found pooled added value for outcome prediction while identifying sample size, reporting and access to individual patient data as priorities for translation \cite{ref30}. The field therefore needs a reproducible mapping from clinical records to the evaluated task: who was included, which channel was counted once, how its target was defined, and which patients informed preprocessing, fitting and threshold selection.

Our evaluation makes the patient this unit of separation and aggregation. The source-held-out analysis then asks a further question about transfer to a different recording population. The interictal analysis illustrates why cohort construction belongs beside the performance table: the original Omni-iEEG benchmark already uses subject-level separation, while our filtering and consolidation produce a different set of unique patient--channel pairs \cite{ref24}. Its results describe that explicitly defined population. Agreement on a dataset name alone does not align inclusion rules, label semantics or endpoints.

Shared data formats such as iEEG-BIDS support this process \cite{ref31}. Reproducible comparisons additionally require split identifiers, channel eligibility rules, preprocessing and threshold procedures, and predictions from the reported runs. This follows the emphasis of TRIPOD+AI on transparent specification of model development and evaluation \cite{ref41}. With these elements fixed, a PR comparison estimates the effect of adding patient context to a particular predictor, and the component ablations identify the contribution of the model's reference operations.

\subsection{Patient Context as a Modeling Variable}
PR addresses a specific part of this inference problem: the reference against which a measured channel is interpreted. Anatomical normative maps show the value of comparing recordings with an explicit distribution of expected physiology \cite{ref21,ref22}. PR uses another source of context, the distribution of the patient's observable channels. Its coordinate transformation allows a shared predictor to operate on the channel's position within that distribution. For a linear readout, Eq. (7) makes the consequence explicit: the same learned coefficients induce patient-dependent coordinate weights and an offset.

The experiments separate the reasons to use this information. Physiological associations identify measurements related to the channel annotations within patients. The half-channel CRPS experiment then tests the reference independently of classification: statistics estimated from one set of channels predict other channels from the same patient. Paired baseline comparisons test whether those relative coordinates help identify the annotated target. This progression connects a distributional property of the recordings to a modeling choice and then to localization performance.

DREChanNet combines this cross-channel context with temporal self-reference. Pre-onset subtraction expresses a channel's change from its own earlier activity, and cross-seizure aggregation retains both its average response and variability. The physiological and raw branches provide explicit measurements and learned waveform representations within this common design. Their joint performance and the PR ablations support using both forms of evidence. The interictal readout applies the same reference principle to separately trained representations, establishing a second application without tying PR to an ictal input sequence.

The reference remains conditional on the implanted set. SEEG implantation is guided by a presurgical hypothesis \cite{ref6,ref7}; changing coverage or the mixture of abnormal and relatively unaffected contacts changes its distribution. PR consequently describes relative position in the observed system. Anatomical normative maps describe departure from an external physiological reference. Combining these two forms of context is a direct extension: anatomy can define the comparison group, while patient statistics can describe its recording-specific location and scale. Spatial perturbation methods for assessing implantation adequacy provide a relevant basis for evaluating such extensions \cite{ref42}.

\subsection{Connecting Localization to Clinical Decisions}
A channel score acquires clinical value through the decision it informs. The ranking experiment provides one concrete use: bringing relevant contacts forward for detailed review. In the displayed frozen run, DREChanNet achieves a Top-5 hit rate of 90.00\%, with 63.00\% precision and 21.98\% recall at five channels (Fig. 4). These endpoints describe both the concentration of labeled EZ channels and the extent of the target covered by a short list. Their joint interpretation connects model output to the amount of evidence a clinician would inspect.

The other experiments address how that evidence becomes available. Additional seizures improve balanced classification in the fixed seizure-sufficiency cohort, supporting aggregation of repeated observations of the same implanted system. The interictal application supplies a complementary route when the available recording state differs. Clinical iEEG interpretation already integrates ictal and interictal observations \cite{ref7}; a reusable reference operation can help organize these observations within their respective models. The source-held-out results further evaluate whether the resulting predictor transfers across the contributing populations.

The next assessment should measure how this channel evidence changes clinical work. A prospective study can present a fixed model's ranked contacts and corresponding physiological context during multidisciplinary review, then record review time, changes in the localization hypothesis, requests for additional evidence and agreement among assessors. Seizure outcome and preservation of function would assess the downstream value of the resulting decisions. DECIDE-AI provides a framework for examining both system performance and the human factors of this early clinical use \cite{ref43}. Including different pathologies, implantation strategies and postoperative outcomes would connect the defined channel task to the broader surgical population. This places the contribution of DREChanNet and PR at a useful point in the clinical chain: organizing patient-specific electrophysiological evidence so that its interpretation and consequences can be evaluated explicitly.

\section{Conclusion}
DREChanNet combines physiological descriptors and learned temporal representations for ictal channel localization, with PR processing linking both branches to the patient's observable channel distribution. The accompanying plug-in adds the same form of context to existing predictors and extends to an interictal readout. Feature associations, held-out-channel reference prediction, paired comparisons and component ablations support this design. By defining the patient-level task alongside the reference operation, the study connects a practical modeling contribution to the broader challenge of translating heterogeneous iEEG observations into interpretable localization evidence.

\bibliographystyle{unsrtnat}
\bibliography{references}
\flushcolsend
\clearpage
\includepdf[pages=-,fitpaper,addtotoc={1,section,1,Supplementary methods,supp:methods}]{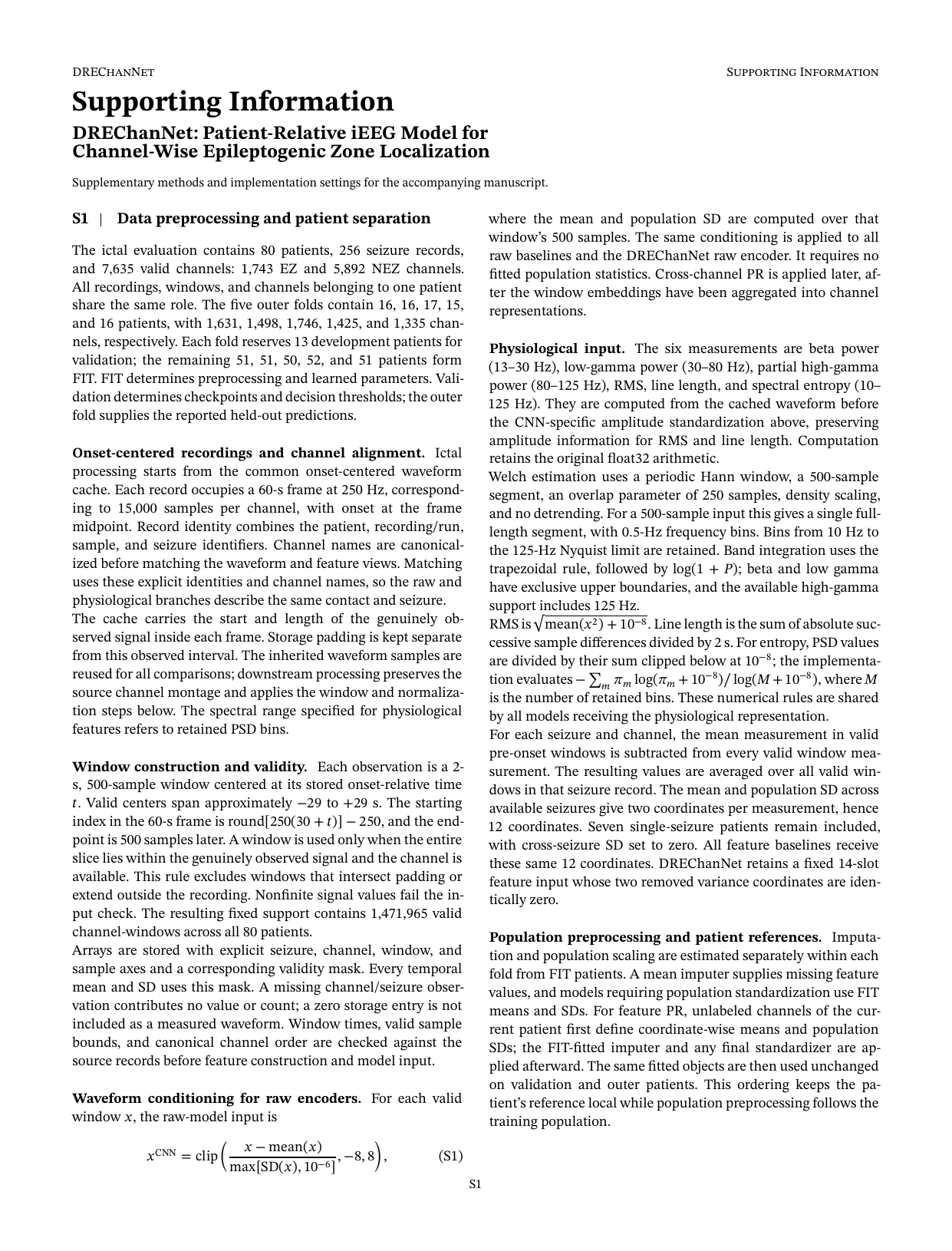}
\end{document}